\documentclass[%
 reprint,
 amsmath,amssymb,
 aps,
 superscriptaddress,
]{revtex4-2}

\usepackage{graphicx}
\usepackage{dcolumn}
\usepackage{bm}
\usepackage{physics}
\usepackage{ulem}
\usepackage[colorlinks=true,linkcolor=black,anchorcolor=black,citecolor=black,filecolor=black,menucolor=black,runcolor=black,urlcolor=black]{hyperref}

\begin{document}

\preprint{APS/123-QED}

\title{Magneto-Acoustic Waves in antiferromagnetic CuMnAs excited by Surface Acoustic Waves}
\title{Néel vector waves in antiferromagnetic CuMnAs excited by Surface Acoustic Waves}

\author{M. Waqas Khaliq}
\email{mkhaliq@cells.es}
\affiliation{Dept. of Condensed Matter Physics, University of Barcelona, 08028-Barcelona, Spain}
\affiliation{ALBA Synchrotron Light Facility, 08290-Cerdanyola del Valles, Barcelona, Spain}
\author{Oliver Amin}
\affiliation{School of Physics and Astronomy, University of Nottingham, NG7 2RD, UK}
\author{Alberto Hernández-Mínguez}
\affiliation{Paul Drude Institute for Solid State Electronics, 10117-Berlin, Germany}
\author{Marc Rovirola}
\affiliation{Dept. of Condensed Matter Physics, University of Barcelona, 08028-Barcelona, Spain}
\affiliation{Institute of Nanoscience and Nanotechnology (IN2UB), University of Barcelona, 08028-Barcelona, Spain}
\author{Blai Casals}
\affiliation{Institute of Nanoscience and Nanotechnology (IN2UB), University of Barcelona, 08028-Barcelona, Spain}
\affiliation{Dept. of Applied Physics, University of Barcelona, 08028-Barcelona, Spain}
\author{Khalid Omari}
\affiliation{Dept. of Electronic Engineering, Royal Holloway, University of London, TW20 0EX-Egham, UK}
\author{Sandra Ruiz-Gómez}
\affiliation{Max Planck Institute for Chemical Physics of Solids, 01187-Dresden, Germany}
\author{Simone Finizio}
\affiliation{Swiss Light Source, Paul Scherrer Institute, 5232-Villigen, Switzerland}
\author{Richard P. Campion}
\affiliation{School of Physics and Astronomy, University of Nottingham, NG7 2RD, UK}
\author{Kevin W. Edmonds}
\affiliation{School of Physics and Astronomy, University of Nottingham, NG7 2RD, UK}
\author{Vít Novák}
\affiliation{Institute of Physics AS CR, 16253-Prague, Czech Republic}
\author{Anna Mandziak}
\affiliation{SOLARIS National Synchrotron Radiation Centre, 30-392-Kraków, Poland}
\author{Lucia Aballe}
\affiliation{ALBA Synchrotron Light Facility, 08290-Cerdanyola del Valles, Barcelona, Spain}
\author{Miguel Angel Niño}
\affiliation{ALBA Synchrotron Light Facility, 08290-Cerdanyola del Valles, Barcelona, Spain}
\author{Joan Manel Hernàndez}
\affiliation{Dept. of Condensed Matter Physics, University of Barcelona, 08028-Barcelona, Spain}
\affiliation{Institute of Nanoscience and Nanotechnology (IN2UB), University of Barcelona, 08028-Barcelona, Spain}
\author{Peter Wadley}
\affiliation{School of Physics and Astronomy, University of Nottingham, NG7 2RD, UK}
\author{Ferran Macià}
\email{ferran.macia@ub.edu}
\affiliation{Dept. of Condensed Matter Physics, University of Barcelona, 08028-Barcelona, Spain}
\affiliation{Institute of Nanoscience and Nanotechnology (IN2UB), University of Barcelona, 08028-Barcelona, Spain}
\author{Michael Foerster}
\email{mfoerster@cells.es}
\affiliation{ALBA Synchrotron Light Facility, 08290-Cerdanyola del Valles, Barcelona, Spain}
\date{\today}

\begin{abstract}
Magnetoelastic effects in antiferromagnetic CuMnAs are investigated by applying dynamic strain in the 0.01\% range through surface acoustic waves in the GaAs substrate. The magnetic state of the CuMnAs/GaAs is characterized by a multitude of submicron-sized domains which we image by x-ray magnetic linear dichroism combined with photoemission electron microscopy. Within the explored strain range, CuMnAs shows magnetoelastic effects in the form of Néel vector waves with micrometer wavelength, which corresponds to an averaged overall spin-axis rotation up to 2.4$^\circ$ driven by the time-dependent strain from the surface acoustic wave. Measurements at different temperatures indicate a reduction of the wave amplitude when lowering the temperature. However, no domain wall motion has been detected on the nanosecond timescale.         
\end{abstract}

\maketitle



Antiferromagnets (AFM) have become a focus of recent research in spintronics, mostly thanks to their potential advantages for future devices. Their low stray fields and robustness versus external magnetic fields are favorable for the further down-scaling of memory elements and their high-frequency internal resonances promise higher intrinsic speed limits for operation. However, together with these advantages also challenges arise, for example, related with the readout and mostly the writing process. Magnetic field control, although not completely impossible \cite{sapozhnik2018direct}, is impractical due to the field magnitudes required ($\gtrsim$ 2 T) in order to overcome exchange energy and modify the magnetization of the two sublattices existing in the AFM. Domain modification by electrical currents has been demonstrated through Spin transfer/orbit torque \cite{wadley2016electrical} as well as through thermoelastic effects \cite{meer2021direct}. Specifically for CuMnAs, the manipulation of antiferromagnetic domains in thin films has been studied by means of injecting current pulses \cite{wadley2016electrical,wadley2018current,janda2020magneto} and defects \cite{reimers2022defect}.  Other approaches use the transitions to a ferromagnetic (FM) phase, like in FeRh \cite{fina2020strain}, or the coupling with a FM \cite{wang2021perpendicular}, which compromise many of the potential advantages (stray fields, speed) of AFM materials for usage in a real device. On the other hand the appearance of closure domain-like features in patterned AFM samples has been attributed to magnetoelastic effects caused by shape dependent strain \cite{meer2022strain}, which suggests that much smaller energies than the exchange energy may be enough to rotate the Néel vector.

Surface acoustic waves (SAW), are propagating elastic deformations in the upper micrometric layer of a crystal. SAW can be conveniently excited in piezoelectric materials by radiofrequency electrical signals applied to an antenna-like structure named interdigitated transducer (IDT). Typical strain amplitudes achieved under UHV conditions can reach the range of $2\times10^{-4}$ for LiNbO$_3$ in the hundreds of MHz to GHz frequency regime \cite{foerster2019quantification}. There is a sizable interaction of SAW with FM systems in heterostructures, which is driven by the transfer of the time dependent strain state of the underlayer/substrate into the FM overlayer. The interaction is mediated by the magnetoelastic effect and has been investigated by a growing number of groups \cite{Hernandez2006,davis2010magnetization,weiler2011elastically,weiler2012spin,thevenard2016precessional,Labanowski2016,foerster2017direct,Kuszewski_2018,MRS2018,adhikari2021surface,casals2020generation,mueller2022imaging,seemann2022magnetoelastic} (see review articles \cite{SAW_roadmap_2019,yang2021acoustic,puebla2022perspectives} and references therein). GaAs, apart from being a suitable substrate for epitaxial CuMnAs growth, has substantial applications in optoelectronic devices due to its outstanding photovoltaic properties \cite{dong2019generation} and robust piezoelectricity \cite{rampal2021optical}.

\begin{figure*}[htb]
    \centering
    \includegraphics[width=0.9\textwidth]{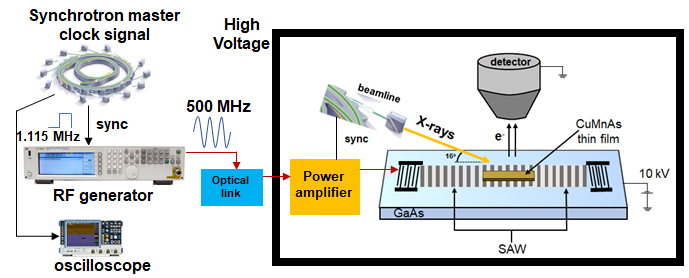}
    \caption{
    \textbf Schematic illustration of the stroboscopic experiment. The CuMnAs thin film is analyzed in the PEEM microscope. SAWs are generated in the GaAs substrate by applying an electrical signal which is synchronized with the synchrotron frequency (repetition rate time of the X-ray pulses). The PEEM image is formed by photoelectrons emitted from the sample under the X-ray illumination.  
    }
    \label{scheme}
\end{figure*}

In this paper we generate Néel vector waves in collinear antiferromagnet CuMnAs induced by the time dependent strain from the supporting GaAs substrate. We obtain direct images using stroboscopic X-ray magnetic linear dichroism combined with photoemission electron microscopy (XMLD-PEEM) of both dynamic strain and Néel vector oscillations with a quantification of the amplitude of the spin axis rotation up to 2.4$^\circ$. The overall amplitude of the observed Néel vector oscillations decreased with lowering temperature. \\


Experiments were carried out at the CIRCE beamline of the ALBA Synchrotron light source \cite{aballe2015alba}. The CuMnAs epitaxial thin films with 45 nm thickness were grown on lattice-matched GaAs substrate by molecular beam epitaxy as described previously \cite{Wadley2013}. To generate SAWs with a frequency tuned to the synchrotron repetition rate (500 MHz), IDT with a finger periodicity of 5.73 $\mu$m (which determines the SAW wavelength) were patterned and deposited on the GaAs by electron beam lithography and metal evaporation. The sample was mounted on a printed circuit board (PCB) inside the sample holder and the IDT were contacted with wire bonds to apply electrical signals. The schematic illustration of the experiment is presented in Fig. \ref{scheme}. The radiofrequency signal applied to the IDT generates a SAW beam \cite{von2020promotion} traveling along the [110] crystalline direction of the GaAs substrate and confined to a depth in the order of the SAW wavelength \cite{puebla2022perspectives}. The SAW causes a periodic in-plane, parallel to the SAW propagation direction, and out-of-plane change in the substrate lattice constant which is transferred as strain to the CuMnAs film. 

\begin{figure}[ht]
    \centering
    \includegraphics[width=1\columnwidth]{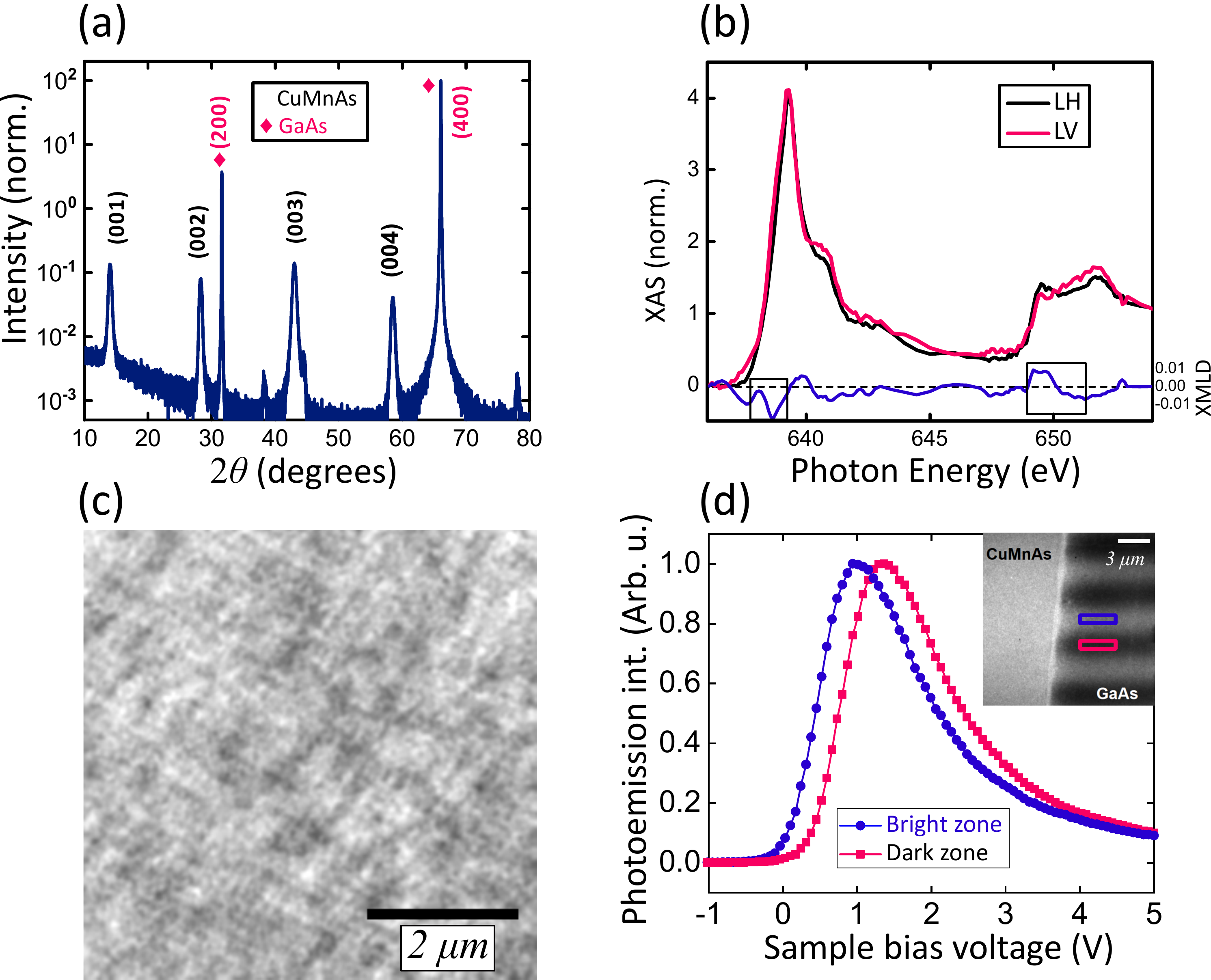}
    \caption{
    \textbf{(a)} XRD measurements of 45 nm thin film of CuMnAs on GaAs showing the film (black) and substrate (red) peaks \textbf{b)} XAS and XMLD spectra of CuMnAs/GaAs at the Mn $L_{3,2}$ edges, obtained with horizontal (black) and vertical (red) linear polarization. \textbf{c)} An XMLD image of CuMnAs thin film at $T\simeq 223$ K by employing linear horizontal polarization component without applying SAW signal. \textbf{d)} Quantification of the SAW: Average intensity (number of emitted electrons) from the red and blue rectangles in the inset as a function of sample bias voltage. The inset shows the XPEEM image of the SAW in GaAs at 0.7 V bias.  }
    \label{fig2}
\end{figure}

In order to assess the structure of the CuMnAs thin film, X-ray diffraction (XRD) measurements were carried out on the sample using a laboratory diffractometer before the synchrotron experiment. The XRD pattern in Fig.\ \ref{fig2}a shows the peaks of both the CuMnAs film (black) and the GaAs substrate (red) with respect to $2\theta$. The tetragonal crystal system (P4/nmm) and the planes (001), (002), (003), and (004) of the antiferromagnetic thin film were identified with the JCDPS-ICDD 01-082-3986 \cite{nateprov2011structure,hills2015paramagnetic}.

X-ray absorption spectroscopy (XAS) measurements of the CuMnAs were performed detecting low energy secondary electrons in the PEEM while scanning the photon energy. The beamline is equipped with an undulator that enables the control of the incoming X-ray polarization, for example between linear horizontal (electric field vector in the sample plane) and vertical (electric field vector under 16 degree to the sample normal) polarization directions. X-ray absorption spectra at the Mn $L_{3,2}$ edges, with both polarizations, and their difference (linear dichroism, XMLD), are depicted in Fig.\ \ref{fig2}b at $T\simeq 235$ K. The features of the XMLD spectrum (blue line), marked by black boxes, are similar to previously published data \cite{wadley2015antiferromagnetic}.

The imaging of antiferromagnetic domains in the CuMnAs thin film is performed by PEEM employing XMLD contrast. Such a contrast is obtained by subtracting different images taken with linear horizontal polarization at energies before and at the $L_3$ absorption peak (See details of the methodology in Supp. Mater. I). Fig.\ \ref{fig2}c shows the domains arrangement in the film at 223 K nominal temperature without any SAW applied. Equivalent images taken with linear vertical polarization (electric field vector 16$^\circ$ to the sample normal) did not show visible contrast, confirming a dominant in-plane Néel vector.

In order to quantify the dynamic magnetoelastic effects in the CuMnAs film, it is necessary to determine first the amplitude of the SAW-induced strain in the GaAs substrate. The inset of Fig.\ \ref{fig2}d depicts an XPEEM image of the sample surface, measured with 500 MHz pulsed synchrotron X-rays. An intensity contrast with a periodicity matching the SAW wavelength, i.e., 5.73 $\mu$m, is evident in the sample region not covered by the CuMnAs film. This contrast originates from the oscillating piezoelectric potential accompanying the strain wave at the surface of the GaAs substrate. The photoelectron spectra displayed in Fig. \ref{fig2}d show the average intensity (number of detected electrons) at the areas marked by the red and blue rectangles in the inset image, recorded as a function of the bias voltage applied to the substrate for a fixed energy analyzer configuration. The voltage shift between these curves, obtained by selecting the positions of maximum slope in both spectra, amounts to 0.35 V and corresponds to the peak-to-peak amplitude of the oscillating piezoelectric potential. This value is used to calculate the amplitude of the strain field by numerically solving the coupled differential equations of the mechanical and electrical displacement, obtaining values in the range of 0.01\% at the sample surface. Details on stroboscopic XPEEM measurements with synchronized SAW can be found in \cite{foerster2019quantification}.\\

\begin{figure}[ht]
    \centering
    \includegraphics[width=\columnwidth]{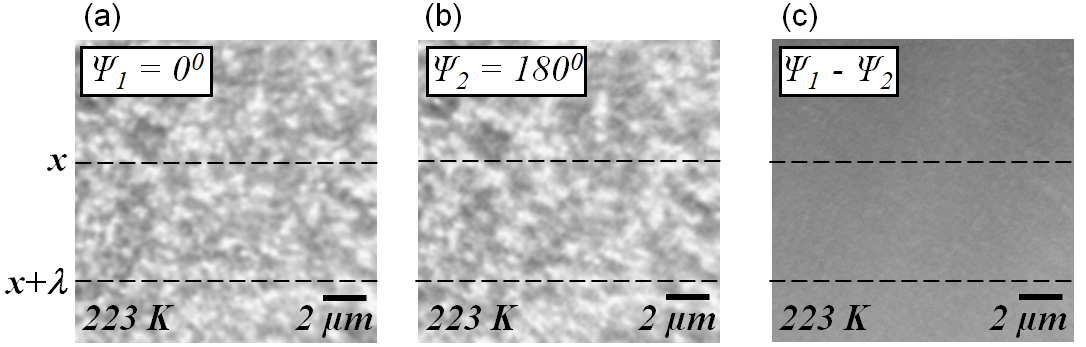}
    \caption{XMLD images of CuMnAs with opposite electronic phases $\psi = 0$$^\circ$ in \textbf{a)} and $\psi=180$$^\circ$ in \textbf{b)} of the SAW excitation at $T \simeq 223$ K. The direction of SAW propagation is perpendicular to the dashed lines which are separated by one wavelength of 5.73 $\mu$m. \textbf{c)} Image obtained by subtracting the images at opposite phases, at the same contrast scale. No evident variations in the domain boundaries can be observed.
    }
    \label{fig3}
\end{figure}

Fig. 3 shows XMLD images taken while the SAW is applied. The bright and dark areas in Fig.\ \ref{fig3} correspond to domains with spin axis parallel and perpendicular to the x-ray polarization, with a typical domain size below one micrometer. The presence of domains with continuously differing gray scale contrast indicates the absence of significant in-plane anisotropy for the spin axis in the sample. Between Fig.\ \ref{fig3}a and \ref{fig3}b, the phase of the radiofrequency signal exciting the SAW was shifted by 180$^\circ$. Thus, the phase of the SAW in any given position is inverted for the stroboscopic measurement, i.e., when the X-rays hit the sample. A close inspection of the individual images shown in Figs. \ref{fig3}a and \ref{fig3}b as well the difference image (Fig. \ref{fig3}c), same gray scale, does not show any observable change of the domain boundaries. The domain wall motion in the CuMnAs in the present experiment is thus either negligible or below the detection limit. The difference images are used to eliminate the static domain contrast and enhance the dynamic changes induced by the SAW. 
\begin{figure}[ht]
    \centering
    \includegraphics[width=\columnwidth]{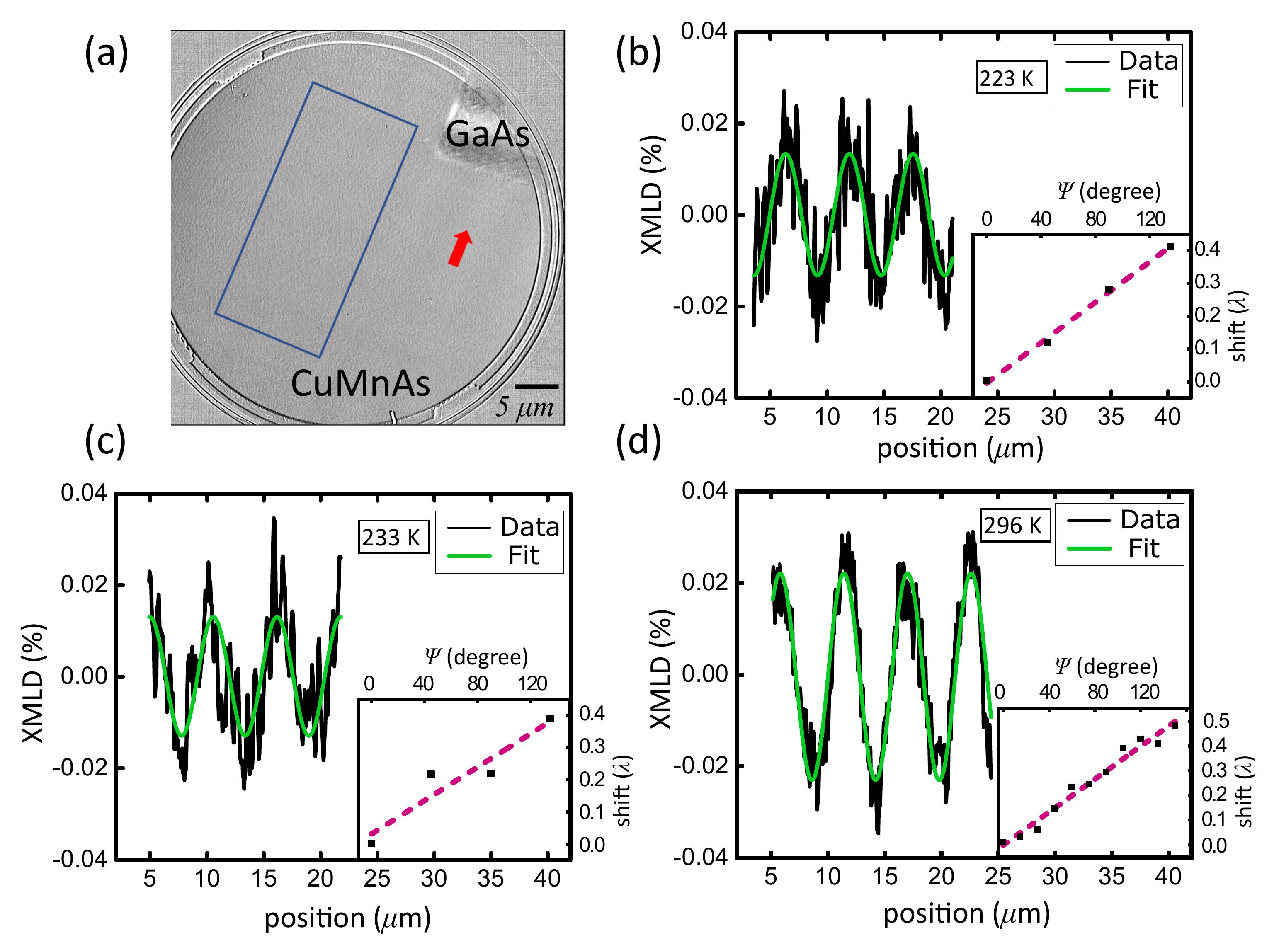}
    \caption{Néel vector wave (spin axis rotation wave) in CuMnAs observed by XMLD-PEEM. \textbf{a)} example of an XMLD difference image from opposite SAW phases. Line profiles are taken along the blue rectangle box, in the propagation direction of the SAW (red arrow). \textbf{(b-d)} XMLD signal (in \%) along the line profile after averaging at $T \simeq  223$ K in \textbf{b)}, $T \simeq  233$ K, in \textbf{c)} and $T \simeq  296$ K in \textbf{d)}. The black line in each plot shows the data and the green line is the sinusoidal function curve fitted to the data points. (insets) fit results for single profiles before averaging: fitted phase shift as a function of the shift $\psi$ in the electrical excitation.
    }
    \label{fig4}
\end{figure}

Several difference images equivalent to the one shown in Fig. 3c, were recorded while changing the electronic phase by a small amount (typically 15$^\circ$). Broad line profiles along the direction of the SAW (blue box in Fig.\ \ \ref{fig4}a) were extracted. For each line profile a background correction is performed by subtracting the signal averaged over exactly one wavelength in order to highlight the oscillatory component(s). Please refer to Supp. Mater. II for details on the data analysis. The line profiles obtained for all phases are then averaged, after shifting them to account for the corresponding phase difference of the SAW from the electronic signal. The results are plotted in Fig.\ \ref{fig4}b–d and correspond to different data sets at $T=223$ K, $T=233$ K, and $T=296$ K, respectively. Green lines are best fits with sinusoidal functions, which is used to obtain the amplitude of the Néel vector (spin axis rotation) signal. In the insets of Fig.\ \ref{fig4}b–d, we show as a validation the result for each single profile, i.e., the fitted experimental phase shift in units of the wavelength $\lambda$. These values are not used for further analysis, but their excellent agreement with the electronic phase $\psi$ applied to the IDT clearly demonstrates that the Néel vector oscillations are driven by SAW.\\




\begin{table*}[ht]
  \centering
  \begin{tabular}{|*{7}{c|}}
    \hline
     Temperature & Max. XMLD & SAW orientation & Strain & XMLD wave & Néel vector & Efficiency \\
      & Domain Contrast & w.r.t. X-rays & Amplitude & amplitude & rotation & \\
      (K) & ($\times 10^{-3}$) & (degrees) & ($\times 10^{-4}$) & ($\times 10^{-4}$) & (degrees) & (degrees/strain $\times 10^{-4}$) \\
    \hline
    223 & $7.6\pm1.0$ & 65 & $0.86\pm 0.05$ & 1.33 & $1.56\pm 0.21$ & $1.81\pm 0.27$ \\
    \hline
     233 & $6.4\pm 1.0$ & 65 & $0.86\pm 0.05$ & 1.29 & $1.80\pm 0.29$ & $2.09\pm 0.36$ \\
    \hline
     296 & $5.3\pm 0.7$ & 90 & $0.75\pm 0.05$ & 2.26 & $2.44\pm 0.33$ & $3.26\pm 0.49$ \\
    \hline
  \end{tabular}
  \caption{Quantitative results of the Neel vector wave in CuMnAs at different temperatures. }
  \label{tab}
\end{table*}

Now we turn to the quantitative analysis of the Néel vector rotation amplitudes for the three data sets at different temperatures, i.e.,  $T=223$ K, $T=233$ K, and $T=296$ K. The results are summarized in Table\ \ref{tab}. Due to experimental constraints, the low-temperature data was taken with an angle of 65$^\circ$ between SAW and probing X-rays, while room temperature data was taken with a 90$^\circ$ angle. The conversion of XMLD amplitude to rotation in degrees as well as the correction factor for the reduced sensitivity under 65$^\circ$ is calculated in Supp. Mater. III.  As mentioned above, the applied strain from the SAW has been calculated from the voltage shift of the secondary electron spectra detected in the PEEM (Fig.\ \ref{fig2}d). The typical strain amplitudes achieved in our experiments on GaAs are ($0.75-1.5)\times10^{-4}$. The XMLD wave amplitude was determined from fits to the averaged line profiles in Fig.\ \ref{fig4}b–d. In order to convert those numbers into the corresponding rotation of the spin axis, we take first into account the temperature dependence of the XMLD contrast in CuMnAs. The XMLD wave amplitude is thus normalized to the maximum XMLD contrast for each temperature in the static domain image (See, Fig.\ \ref{fig2}c for example), taken as average in three different locations each. We then consider the film to be populated by an equal portion of domains in all direction, i.e., without net in-plane anisotropy and calculate for each domain \emph{i)} the rotation angle as fraction of a maximum angle $\phi_0$ as explained in Supp. Mater. III and \emph{ii)} the sensitivity of the XMLD signal to this rotation as function of the domain spin axis.  

\begin{figure}[ht]
    \centering
    \includegraphics[width=1\columnwidth]{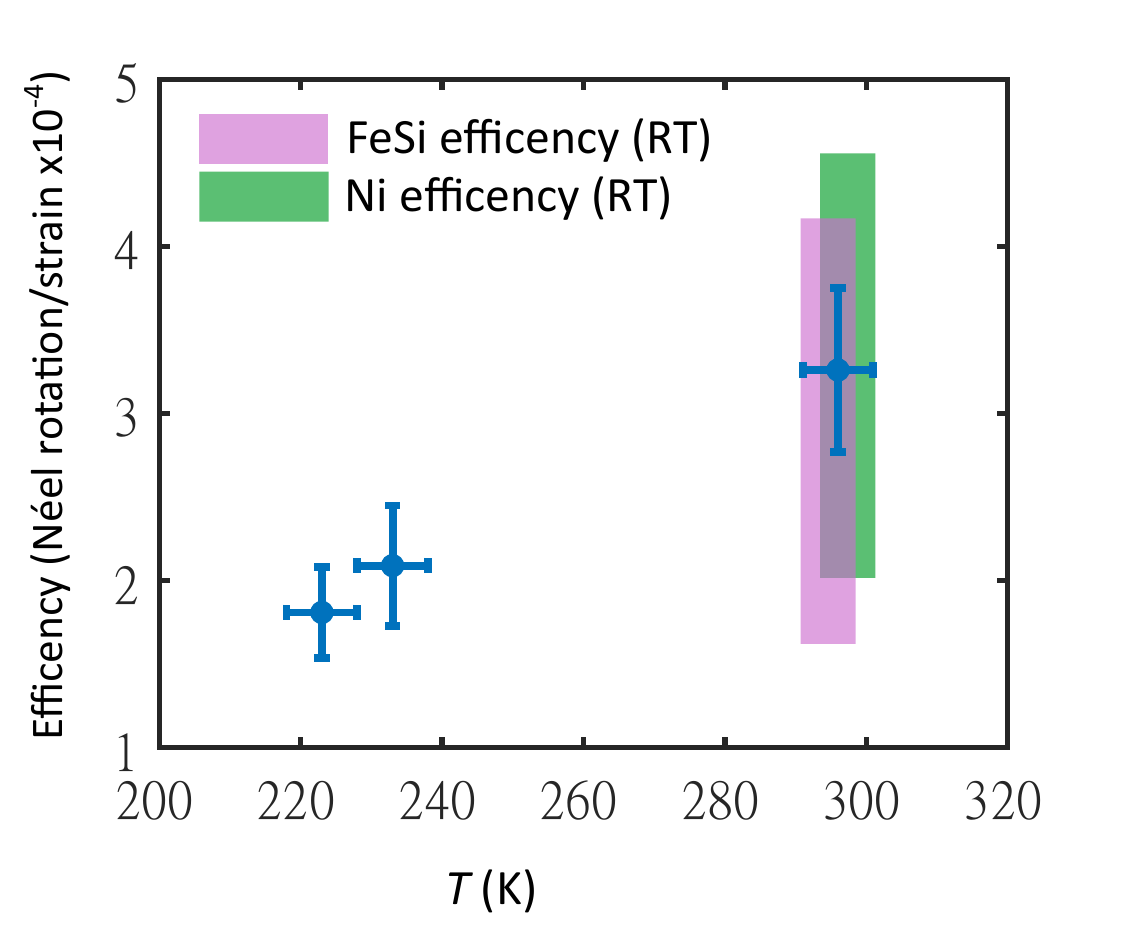}
    \caption{Efficiency of the magnetoacoustic wave excitation in CuMnAs as a function of temperature in comparison with all data at 500 MHz SAW. The area for Ferromagnetic samples corresponds to Ni and FeSi at room temperature (RT) and covers the range between zero external field and resonance.}
    \label{fig5}
\end{figure}

The numbers reported in Table\ \ref{tab} show that the spin axis rotation wave driven by SAW in CuMnAs can reach a sizable 2.44$^\circ$ at room temperature. We plotted in Fig.\ \ref{fig5} the efficiency of the SAW induced Néel vector wave defined as the overall variation divided by the SAW strain for the three temperatures. A similar quantity is added into the graph for magnetoacoustic waves in ferromagnetic samples measured through XMCD \cite{casals2020generation,rovirola2022resonant}. Values for ferromagnetic Ni \cite{casals2020generation} and Heusler alloy {{Fe$_3$Si}} \cite{rovirola2022resonant} are shown as broad bands, because they depend on the external applied magnetic field, showing a resonance like peak (Nickel showed an efficency of 2 to 4.5 and FeSi from 1.6 to 4.1). These results indicate a sizable dynamic magneto-elastic effect in CuMnAs induced by SAW and a comparable efficiency like in FM materials. We notice here that detection of spin axis rotation in CuMnAs is more challenging compared to magnetoacoustic waves in ferromagnets mainly because the XMLD contrast is weaker and because the sample is in a multi-domain state.


A reduction of the Néel vector wave excitation efficiency when lowering temperature, i.e., from 3.3$^\circ$/$10^{-4}$ strain at room temperature (296 K) to 1.8$^\circ$/$10^{-4}$ at 223 K has been observed. While this apparent reduction needs further experiments to verify \cite{grigorev2022opticallytriggered,lyons2023acoustically}, it may indicate an increase of the energy barrier for the spin axis rotation as temperature lowers (a type of “freezing”).\\

In summary, we have investigated high frequency (500 MHz) magnetoelastic effects in antiferromagnetic CuMnAs excited by SAW in the GaAs substrate using XMLD-PEEM. An averaged magneto-acoustic wave signal can be detected in XMLD, corresponding to a rotation of the spin axis in the individual domains by up to $\pm2.4^\circ$. The efficiency of the Néel vector excitation in CuMnAs is a proof that magnetoelastic effects are a viable way to manipulate antiferromagnetic systems, even on the subnanosecond time scale. Moreover, in static condition, the CuMnAs thin film is characterized by a multidomain configuration of submicron size. No SAW-induced motion of domain walls has been detected, which could be related to intrinsic pinning due the film microstructure. For the future, a combination of GaP substrates with ZnO based IDT can enable to study SAW driven effects in high quality single domain patterns of CuMnAs.

\begin{acknowledgments}
The authors are thankful to Spanish Ministry of Science, Innovation, Universities, and DOC-FAM who have received funding from the European Union's Horizon 2020 research and innovation program under the Marie Sklodowska-Curie grant agreement No. 754397. MWK MF, MAN, and LA acknowledge the funding from MICINN through grant numbers RTI2018-095303-B-C53 and PID2021-122980OB-C54. FM, MR, BC, and JMH are grateful to funding from MCIN/AEI/10.13039/501100011033 through grant number: PID2020-113024GB-100. OA, KO, KWE, RPC, and PW acknowledge funding from EU FET Open RIA Grant no 766566. VN is grateful to MEYS Grant No. LM2018110. This work has also been supported by the ALBA inHouse Research Program.
\end{acknowledgments}







\begin{thebibliography}{37}%
	\makeatletter
	\providecommand \@ifxundefined [1]{%
		\@ifx{#1\undefined}
	}%
	\providecommand \@ifnum [1]{%
		\ifnum #1\expandafter \@firstoftwo
		\else \expandafter \@secondoftwo
		\fi
	}%
	\providecommand \@ifx [1]{%
		\ifx #1\expandafter \@firstoftwo
		\else \expandafter \@secondoftwo
		\fi
	}%
	\providecommand \natexlab [1]{#1}%
	\providecommand \enquote  [1]{``#1''}%
	\providecommand \bibnamefont  [1]{#1}%
	\providecommand \bibfnamefont [1]{#1}%
	\providecommand \citenamefont [1]{#1}%
	\providecommand \href@noop [0]{\@secondoftwo}%
	\providecommand \href [0]{\begingroup \@sanitize@url \@href}%
	\providecommand \@href[1]{\@@startlink{#1}\@@href}%
	\providecommand \@@href[1]{\endgroup#1\@@endlink}%
	\providecommand \@sanitize@url [0]{\catcode `\\12\catcode `\$12\catcode
		`\&12\catcode `\#12\catcode `\^12\catcode `\_12\catcode `\%12\relax}%
	\providecommand \@@startlink[1]{}%
	\providecommand \@@endlink[0]{}%
	\providecommand \url  [0]{\begingroup\@sanitize@url \@url }%
	\providecommand \@url [1]{\endgroup\@href {#1}{\urlprefix }}%
	\providecommand \urlprefix  [0]{URL }%
	\providecommand \Eprint [0]{\href }%
	\providecommand \doibase [0]{https://doi.org/}%
	\providecommand \selectlanguage [0]{\@gobble}%
	\providecommand \bibinfo  [0]{\@secondoftwo}%
	\providecommand \bibfield  [0]{\@secondoftwo}%
	\providecommand \translation [1]{[#1]}%
	\providecommand \BibitemOpen [0]{}%
	\providecommand \bibitemStop [0]{}%
	\providecommand \bibitemNoStop [0]{.\EOS\space}%
	\providecommand \EOS [0]{\spacefactor3000\relax}%
	\providecommand \BibitemShut  [1]{\csname bibitem#1\endcsname}%
	\let\auto@bib@innerbib\@empty
	\bibitem [{\citenamefont {Sapozhnik}\ \emph {et~al.}(2018)\citenamefont
		{Sapozhnik}, \citenamefont {Filianina}, \citenamefont {Bodnar}, \citenamefont
		{Lamirand}, \citenamefont {Mawass}, \citenamefont {Skourski}, \citenamefont
		{Elmers}, \citenamefont {Zabel}, \citenamefont {Kl{\"a}ui},\ and\
		\citenamefont {Jourdan}}]{sapozhnik2018direct}%
	\BibitemOpen
	\bibfield  {author} {\bibinfo {author} {\bibfnamefont {A.~A.}\ \bibnamefont
			{Sapozhnik}}, \bibinfo {author} {\bibfnamefont {M.}~\bibnamefont
			{Filianina}}, \bibinfo {author} {\bibfnamefont {S.~Y.}\ \bibnamefont
			{Bodnar}}, \bibinfo {author} {\bibfnamefont {A.}~\bibnamefont {Lamirand}},
		\bibinfo {author} {\bibfnamefont {M.-A.}\ \bibnamefont {Mawass}}, \bibinfo
		{author} {\bibfnamefont {Y.}~\bibnamefont {Skourski}}, \bibinfo {author}
		{\bibfnamefont {H.-J.}\ \bibnamefont {Elmers}}, \bibinfo {author}
		{\bibfnamefont {H.}~\bibnamefont {Zabel}}, \bibinfo {author} {\bibfnamefont
			{M.}~\bibnamefont {Kl{\"a}ui}},\ and\ \bibinfo {author} {\bibfnamefont
			{M.}~\bibnamefont {Jourdan}},\ }\bibfield  {title} {\bibinfo {title} {Direct
			imaging of antiferromagnetic domains in $\text{Mn}_2\text{Au}$ manipulated by
			high magnetic fields},\ }\href {https://doi.org/10.1103/PhysRevB.97.134429}
	{\bibfield  {journal} {\bibinfo  {journal} {Physical Review B}\ }\textbf
		{\bibinfo {volume} {97}},\ \bibinfo {pages} {134429} (\bibinfo {year}
		{2018})}\BibitemShut {NoStop}%
	\bibitem [{\citenamefont {Wadley}\ \emph {et~al.}(2016)\citenamefont {Wadley},
		\citenamefont {Howells}, \citenamefont {Å½elezn{\'y}}, \citenamefont
		{Andrews}, \citenamefont {Hills}, \citenamefont {Campion}, \citenamefont
		{Nov{\'a}k}, \citenamefont {Olejnik}, \citenamefont {Maccherozzi},
		\citenamefont {Dhesi} \emph {et~al.}}]{wadley2016electrical}%
	\BibitemOpen
	\bibfield  {author} {\bibinfo {author} {\bibfnamefont {P.}~\bibnamefont
			{Wadley}}, \bibinfo {author} {\bibfnamefont {B.}~\bibnamefont {Howells}},
		\bibinfo {author} {\bibfnamefont {J.}~\bibnamefont {Å½elezn{\'y}}}, \bibinfo
		{author} {\bibfnamefont {C.}~\bibnamefont {Andrews}}, \bibinfo {author}
		{\bibfnamefont {V.}~\bibnamefont {Hills}}, \bibinfo {author} {\bibfnamefont
			{R.~P.}\ \bibnamefont {Campion}}, \bibinfo {author} {\bibfnamefont
			{V.}~\bibnamefont {Nov{\'a}k}}, \bibinfo {author} {\bibfnamefont
			{K.}~\bibnamefont {Olejnik}}, \bibinfo {author} {\bibfnamefont
			{F.}~\bibnamefont {Maccherozzi}}, \bibinfo {author} {\bibfnamefont {S.~S.}\
			\bibnamefont {Dhesi}}, \emph {et~al.},\ }\bibfield  {title} {\bibinfo {title}
		{Electrical switching of an antiferromagnet},\ }\href
	{https://doi.org/10.1126/science.aab1031} {\bibfield  {journal} {\bibinfo
			{journal} {Science}\ }\textbf {\bibinfo {volume} {351}},\ \bibinfo {pages}
		{587} (\bibinfo {year} {2016})}\BibitemShut {NoStop}%
	\bibitem [{\citenamefont {Meer}\ \emph {et~al.}(2021)\citenamefont {Meer},
		\citenamefont {Schreiber}, \citenamefont {Schmitt}, \citenamefont {Ramos},
		\citenamefont {Saitoh}, \citenamefont {Gomonay}, \citenamefont {Sinova},
		\citenamefont {Baldrati},\ and\ \citenamefont {Kl{\"a}ui}}]{meer2021direct}%
	\BibitemOpen
	\bibfield  {author} {\bibinfo {author} {\bibfnamefont {H.}~\bibnamefont
			{Meer}}, \bibinfo {author} {\bibfnamefont {F.}~\bibnamefont {Schreiber}},
		\bibinfo {author} {\bibfnamefont {C.}~\bibnamefont {Schmitt}}, \bibinfo
		{author} {\bibfnamefont {R.}~\bibnamefont {Ramos}}, \bibinfo {author}
		{\bibfnamefont {E.}~\bibnamefont {Saitoh}}, \bibinfo {author} {\bibfnamefont
			{O.}~\bibnamefont {Gomonay}}, \bibinfo {author} {\bibfnamefont
			{J.}~\bibnamefont {Sinova}}, \bibinfo {author} {\bibfnamefont
			{L.}~\bibnamefont {Baldrati}},\ and\ \bibinfo {author} {\bibfnamefont
			{M.}~\bibnamefont {Kl{\"a}ui}},\ }\bibfield  {title} {\bibinfo {title}
		{Direct imaging of current-induced antiferromagnetic switching revealing a
			pure thermomagnetoelastic switching mechanism in $\text{NiO}$},\ }\href
	{https://doi.org/10.1021/acs.nanolett.0c03367} {\bibfield  {journal}
		{\bibinfo  {journal} {Nano Letters}\ }\textbf {\bibinfo {volume} {21}},\
		\bibinfo {pages} {114} (\bibinfo {year} {2021})}\BibitemShut {NoStop}%
	\bibitem [{\citenamefont {Wadley}\ \emph {et~al.}(2018)\citenamefont {Wadley},
		\citenamefont {Reimers}, \citenamefont {Grzybowski}, \citenamefont {Andrews},
		\citenamefont {Wang}, \citenamefont {Chauhan}, \citenamefont {Gallagher},
		\citenamefont {Campion}, \citenamefont {Edmonds}, \citenamefont {Dhesi} \emph
		{et~al.}}]{wadley2018current}%
	\BibitemOpen
	\bibfield  {author} {\bibinfo {author} {\bibfnamefont {P.}~\bibnamefont
			{Wadley}}, \bibinfo {author} {\bibfnamefont {S.}~\bibnamefont {Reimers}},
		\bibinfo {author} {\bibfnamefont {M.~J.}\ \bibnamefont {Grzybowski}},
		\bibinfo {author} {\bibfnamefont {C.}~\bibnamefont {Andrews}}, \bibinfo
		{author} {\bibfnamefont {M.}~\bibnamefont {Wang}}, \bibinfo {author}
		{\bibfnamefont {J.~S.}\ \bibnamefont {Chauhan}}, \bibinfo {author}
		{\bibfnamefont {B.~L.}\ \bibnamefont {Gallagher}}, \bibinfo {author}
		{\bibfnamefont {R.~P.}\ \bibnamefont {Campion}}, \bibinfo {author}
		{\bibfnamefont {K.~W.}\ \bibnamefont {Edmonds}}, \bibinfo {author}
		{\bibfnamefont {S.~S.}\ \bibnamefont {Dhesi}}, \emph {et~al.},\ }\bibfield
	{title} {\bibinfo {title} {Current polarity-dependent manipulation of
			antiferromagnetic domains},\ }\href
	{https://doi.org/10.1038/s41565-018-0079-1} {\bibfield  {journal} {\bibinfo
			{journal} {Nat. Nanotechnol.}\ }\textbf {\bibinfo {volume} {13}},\ \bibinfo
		{pages} {362} (\bibinfo {year} {2018})}\BibitemShut {NoStop}%
	\bibitem [{\citenamefont {Janda}\ \emph {et~al.}(2020)\citenamefont {Janda},
		\citenamefont {Godinho}, \citenamefont {Ostatnicky}, \citenamefont
		{Pfitzner}, \citenamefont {Ulrich}, \citenamefont {Hoehl}, \citenamefont
		{Reimers}, \citenamefont {Soban}, \citenamefont {Metzger}, \citenamefont
		{Reichlova} \emph {et~al.}}]{janda2020magneto}%
	\BibitemOpen
	\bibfield  {author} {\bibinfo {author} {\bibfnamefont {T.}~\bibnamefont
			{Janda}}, \bibinfo {author} {\bibfnamefont {J.}~\bibnamefont {Godinho}},
		\bibinfo {author} {\bibfnamefont {T.}~\bibnamefont {Ostatnicky}}, \bibinfo
		{author} {\bibfnamefont {E.}~\bibnamefont {Pfitzner}}, \bibinfo {author}
		{\bibfnamefont {G.}~\bibnamefont {Ulrich}}, \bibinfo {author} {\bibfnamefont
			{A.}~\bibnamefont {Hoehl}}, \bibinfo {author} {\bibfnamefont
			{S.}~\bibnamefont {Reimers}}, \bibinfo {author} {\bibfnamefont
			{Z.}~\bibnamefont {Soban}}, \bibinfo {author} {\bibfnamefont
			{T.}~\bibnamefont {Metzger}}, \bibinfo {author} {\bibfnamefont
			{H.}~\bibnamefont {Reichlova}}, \emph {et~al.},\ }\bibfield  {title}
	{\bibinfo {title} {Magneto-seebeck microscopy of domain switching in
			collinear antiferromagnet $\text{CuMnAs}$},\ }\href
	{https://doi.org/10.1103/PhysRevMaterials.4.094413} {\bibfield  {journal}
		{\bibinfo  {journal} {Phys. Rev. Mat.}\ }\textbf {\bibinfo {volume} {4}},\
		\bibinfo {pages} {094413} (\bibinfo {year} {2020})}\BibitemShut {NoStop}%
	\bibitem [{\citenamefont {Reimers}\ \emph {et~al.}(2022)\citenamefont
		{Reimers}, \citenamefont {Kriegner}, \citenamefont {Gomonay}, \citenamefont
		{Carbone}, \citenamefont {Krizek}, \citenamefont {Nov{\'a}k}, \citenamefont
		{Campion}, \citenamefont {Maccherozzi}, \citenamefont {Bj{\"o}rling},
		\citenamefont {Amin} \emph {et~al.}}]{reimers2022defect}%
	\BibitemOpen
	\bibfield  {author} {\bibinfo {author} {\bibfnamefont {S.}~\bibnamefont
			{Reimers}}, \bibinfo {author} {\bibfnamefont {D.}~\bibnamefont {Kriegner}},
		\bibinfo {author} {\bibfnamefont {O.}~\bibnamefont {Gomonay}}, \bibinfo
		{author} {\bibfnamefont {D.}~\bibnamefont {Carbone}}, \bibinfo {author}
		{\bibfnamefont {F.}~\bibnamefont {Krizek}}, \bibinfo {author} {\bibfnamefont
			{V.}~\bibnamefont {Nov{\'a}k}}, \bibinfo {author} {\bibfnamefont {R.~P.}\
			\bibnamefont {Campion}}, \bibinfo {author} {\bibfnamefont {F.}~\bibnamefont
			{Maccherozzi}}, \bibinfo {author} {\bibfnamefont {A.}~\bibnamefont
			{Bj{\"o}rling}}, \bibinfo {author} {\bibfnamefont {O.~J.}\ \bibnamefont
			{Amin}}, \emph {et~al.},\ }\bibfield  {title} {\bibinfo {title}
		{Defect-driven antiferromagnetic domain walls in $\text{CuMnAs}$ films},\
	}\href {https://doi.org/10.1038/s41467-022-28311-x} {\bibfield  {journal}
		{\bibinfo  {journal} {Nat. Commun.}\ }\textbf {\bibinfo {volume} {13}},\
		\bibinfo {pages} {724} (\bibinfo {year} {2022})}\BibitemShut {NoStop}%
	\bibitem [{\citenamefont {Fina}\ and\ \citenamefont
		{Fontcuberta}(2020)}]{fina2020strain}%
	\BibitemOpen
	\bibfield  {author} {\bibinfo {author} {\bibfnamefont {I.}~\bibnamefont
			{Fina}}\ and\ \bibinfo {author} {\bibfnamefont {J.}~\bibnamefont
			{Fontcuberta}},\ }\bibfield  {title} {\bibinfo {title} {Strain and voltage
			control of magnetic and electric properties of $\text{FeRh}$ films},\ }\href
	{https://doi.org/10.1088/1361-6463/ab4abd} {\bibfield  {journal} {\bibinfo
			{journal} {J. Phys. D: Appl. Phys.}\ }\textbf {\bibinfo {volume} {53}},\
		\bibinfo {pages} {023002} (\bibinfo {year} {2020})}\BibitemShut {NoStop}%
	\bibitem [{\citenamefont {Wang}\ \emph {et~al.}(2021)\citenamefont {Wang},
		\citenamefont {Hsiao}, \citenamefont {Liao}, \citenamefont {Hsu},
		\citenamefont {Li}, \citenamefont {Hsu}, \citenamefont {Lai}, \citenamefont
		{Tsai}, \citenamefont {Chuang},\ and\ \citenamefont
		{Wei}}]{wang2021perpendicular}%
	\BibitemOpen
	\bibfield  {author} {\bibinfo {author} {\bibfnamefont {B.-Y.}\ \bibnamefont
			{Wang}}, \bibinfo {author} {\bibfnamefont {C.-H.}\ \bibnamefont {Hsiao}},
		\bibinfo {author} {\bibfnamefont {B.-X.}\ \bibnamefont {Liao}}, \bibinfo
		{author} {\bibfnamefont {C.-Y.}\ \bibnamefont {Hsu}}, \bibinfo {author}
		{\bibfnamefont {T.-H.}\ \bibnamefont {Li}}, \bibinfo {author} {\bibfnamefont
			{Y.-L.}\ \bibnamefont {Hsu}}, \bibinfo {author} {\bibfnamefont {Y.-M.}\
			\bibnamefont {Lai}}, \bibinfo {author} {\bibfnamefont {M.-S.}\ \bibnamefont
			{Tsai}}, \bibinfo {author} {\bibfnamefont {T.-H.}\ \bibnamefont {Chuang}},\
		and\ \bibinfo {author} {\bibfnamefont {D.-H.}\ \bibnamefont {Wei}},\
	}\bibfield  {title} {\bibinfo {title} {Perpendicular magnetic anisotropy
			induced by nimn-based antiferromagnetic films with in-plane spin
			orientations: Roles of interfacial and volume antiferromagnetic moments},\
	}\href {https://doi.org/10.1103/PhysRevB.104.024424} {\bibfield  {journal}
		{\bibinfo  {journal} {Physical Review B}\ }\textbf {\bibinfo {volume}
			{104}},\ \bibinfo {pages} {024424} (\bibinfo {year} {2021})}\BibitemShut
	{NoStop}%
	\bibitem [{\citenamefont {Meer}\ \emph {et~al.}(2022)\citenamefont {Meer},
		\citenamefont {Gomonay}, \citenamefont {Schmitt}, \citenamefont {Ramos},
		\citenamefont {Schnitzspan}, \citenamefont {Kronast}, \citenamefont {Mawass},
		\citenamefont {Valencia}, \citenamefont {Saitoh}, \citenamefont {Sinova}
		\emph {et~al.}}]{meer2022strain}%
	\BibitemOpen
	\bibfield  {author} {\bibinfo {author} {\bibfnamefont {H.}~\bibnamefont
			{Meer}}, \bibinfo {author} {\bibfnamefont {O.}~\bibnamefont {Gomonay}},
		\bibinfo {author} {\bibfnamefont {C.}~\bibnamefont {Schmitt}}, \bibinfo
		{author} {\bibfnamefont {R.}~\bibnamefont {Ramos}}, \bibinfo {author}
		{\bibfnamefont {L.}~\bibnamefont {Schnitzspan}}, \bibinfo {author}
		{\bibfnamefont {F.}~\bibnamefont {Kronast}}, \bibinfo {author} {\bibfnamefont
			{M.-A.}\ \bibnamefont {Mawass}}, \bibinfo {author} {\bibfnamefont
			{S.}~\bibnamefont {Valencia}}, \bibinfo {author} {\bibfnamefont
			{E.}~\bibnamefont {Saitoh}}, \bibinfo {author} {\bibfnamefont
			{J.}~\bibnamefont {Sinova}}, \emph {et~al.},\ }\bibfield  {title} {\bibinfo
		{title} {Strain-induced shape anisotropy in antiferromagnetic structures},\
	}\href {https://doi.org/10.1103/PhysRevB.106.094430} {\bibfield  {journal}
		{\bibinfo  {journal} {Physical Review B}\ }\textbf {\bibinfo {volume}
			{106}},\ \bibinfo {pages} {094430} (\bibinfo {year} {2022})}\BibitemShut
	{NoStop}%
	\bibitem [{\citenamefont {Foerster}\ \emph {et~al.}(2019)\citenamefont
		{Foerster}, \citenamefont {Statuto}, \citenamefont {Casals}, \citenamefont
		{Hern{\'a}ndez-M{\'i}nguez}, \citenamefont {Cichelero}, \citenamefont
		{Manshausen}, \citenamefont {Mandziak}, \citenamefont {Aballe}, \citenamefont
		{Hern{\'a}ndez~Ferr{\`a}s},\ and\ \citenamefont
		{Maci{\`a}}}]{foerster2019quantification}%
	\BibitemOpen
	\bibfield  {author} {\bibinfo {author} {\bibfnamefont {M.}~\bibnamefont
			{Foerster}}, \bibinfo {author} {\bibfnamefont {N.}~\bibnamefont {Statuto}},
		\bibinfo {author} {\bibfnamefont {B.}~\bibnamefont {Casals}}, \bibinfo
		{author} {\bibfnamefont {A.}~\bibnamefont {Hern{\'a}ndez-M{\'i}nguez}},
		\bibinfo {author} {\bibfnamefont {R.}~\bibnamefont {Cichelero}}, \bibinfo
		{author} {\bibfnamefont {P.}~\bibnamefont {Manshausen}}, \bibinfo {author}
		{\bibfnamefont {A.}~\bibnamefont {Mandziak}}, \bibinfo {author}
		{\bibfnamefont {L.}~\bibnamefont {Aballe}}, \bibinfo {author} {\bibfnamefont
			{J.}~\bibnamefont {Hern{\'a}ndez~Ferr{\`a}s}},\ and\ \bibinfo {author}
		{\bibfnamefont {F.}~\bibnamefont {Maci{\`a}}},\ }\bibfield  {title} {\bibinfo
		{title} {Quantification of propagating and standing surface acoustic waves by
			stroboscopic {X}-ray photoemission electron microscopy},\ }\href
	{https://doi.org/10.1107/S1600577518015370} {\bibfield  {journal} {\bibinfo
			{journal} {J. Synchrotron Rad.}\ }\textbf {\bibinfo {volume} {26}},\ \bibinfo
		{pages} {184} (\bibinfo {year} {2019})}\BibitemShut {NoStop}%
	\bibitem [{\citenamefont {Hernandez}\ \emph {et~al.}(2006)\citenamefont
		{Hernandez}, \citenamefont {Santos}, \citenamefont {Maci\`a}, \citenamefont
		{Garc\'ia-Santiago},\ and\ \citenamefont {Tejada}}]{Hernandez2006}%
	\BibitemOpen
	\bibfield  {author} {\bibinfo {author} {\bibfnamefont {J.}~\bibnamefont
			{Hernandez}}, \bibinfo {author} {\bibfnamefont {P.}~\bibnamefont {Santos}},
		\bibinfo {author} {\bibfnamefont {F.}~\bibnamefont {Maci\`a}}, \bibinfo
		{author} {\bibfnamefont {A.}~\bibnamefont {Garc\'ia-Santiago}},\ and\
		\bibinfo {author} {\bibfnamefont {J.}~\bibnamefont {Tejada}},\ }\bibfield
	{title} {\bibinfo {title} {Acoustomagnetic pulse experiments in
			{{LiNbO$_3$/Mn$_{12}$}} hybrids},\ }\href {https://doi.org/10.1063/1.2158705}
	{\bibfield  {journal} {\bibinfo  {journal} {Appl. Phys. Lett.}\ }\textbf
		{\bibinfo {volume} {88}},\ \bibinfo {pages} {012503} (\bibinfo {year}
		{2006})}\BibitemShut {NoStop}%
	\bibitem [{\citenamefont {Davis}\ \emph {et~al.}(2010)\citenamefont {Davis},
		\citenamefont {Baruth},\ and\ \citenamefont
		{Adenwalla}}]{davis2010magnetization}%
	\BibitemOpen
	\bibfield  {author} {\bibinfo {author} {\bibfnamefont {S.}~\bibnamefont
			{Davis}}, \bibinfo {author} {\bibfnamefont {A.}~\bibnamefont {Baruth}},\ and\
		\bibinfo {author} {\bibfnamefont {S.}~\bibnamefont {Adenwalla}},\ }\bibfield
	{title} {\bibinfo {title} {Magnetization dynamics triggered by surface
			acoustic waves},\ }\href {https://doi.org/10.1063/1.3521289} {\bibfield
		{journal} {\bibinfo  {journal} {Appl. Phys. Lett.}\ }\textbf {\bibinfo
			{volume} {97}},\ \bibinfo {pages} {232507} (\bibinfo {year}
		{2010})}\BibitemShut {NoStop}%
	\bibitem [{\citenamefont {Weiler}\ \emph {et~al.}(2011)\citenamefont {Weiler},
		\citenamefont {Dreher}, \citenamefont {Heeg}, \citenamefont {Huebl},
		\citenamefont {Gross}, \citenamefont {Brandt},\ and\ \citenamefont
		{Goennenwein}}]{weiler2011elastically}%
	\BibitemOpen
	\bibfield  {author} {\bibinfo {author} {\bibfnamefont {M.}~\bibnamefont
			{Weiler}}, \bibinfo {author} {\bibfnamefont {L.}~\bibnamefont {Dreher}},
		\bibinfo {author} {\bibfnamefont {C.}~\bibnamefont {Heeg}}, \bibinfo {author}
		{\bibfnamefont {H.}~\bibnamefont {Huebl}}, \bibinfo {author} {\bibfnamefont
			{R.}~\bibnamefont {Gross}}, \bibinfo {author} {\bibfnamefont {M.~S.}\
			\bibnamefont {Brandt}},\ and\ \bibinfo {author} {\bibfnamefont {S.~T.}\
			\bibnamefont {Goennenwein}},\ }\bibfield  {title} {\bibinfo {title}
		{Elastically driven ferromagnetic resonance in nickel thin films},\ }\href
	{https://doi.org/10.1103/PhysRevLett.106.117601} {\bibfield  {journal}
		{\bibinfo  {journal} {Physical Review Letters}\ }\textbf {\bibinfo {volume}
			{106}},\ \bibinfo {pages} {117601} (\bibinfo {year} {2011})}\BibitemShut
	{NoStop}%
	\bibitem [{\citenamefont {Weiler}\ \emph {et~al.}(2012)\citenamefont {Weiler},
		\citenamefont {Huebl}, \citenamefont {Goerg}, \citenamefont {Czeschka},
		\citenamefont {Gross},\ and\ \citenamefont {Goennenwein}}]{weiler2012spin}%
	\BibitemOpen
	\bibfield  {author} {\bibinfo {author} {\bibfnamefont {M.}~\bibnamefont
			{Weiler}}, \bibinfo {author} {\bibfnamefont {H.}~\bibnamefont {Huebl}},
		\bibinfo {author} {\bibfnamefont {F.~S.}\ \bibnamefont {Goerg}}, \bibinfo
		{author} {\bibfnamefont {F.~D.}\ \bibnamefont {Czeschka}}, \bibinfo {author}
		{\bibfnamefont {R.}~\bibnamefont {Gross}},\ and\ \bibinfo {author}
		{\bibfnamefont {S.~T.}\ \bibnamefont {Goennenwein}},\ }\bibfield  {title}
	{\bibinfo {title} {Spin pumping with coherent elastic waves},\ }\href
	{https://doi.org/10.1103/PhysRevLett.108.176601} {\bibfield  {journal}
		{\bibinfo  {journal} {Physical Review Letters}\ }\textbf {\bibinfo {volume}
			{108}},\ \bibinfo {pages} {176601} (\bibinfo {year} {2012})}\BibitemShut
	{NoStop}%
	\bibitem [{\citenamefont {Thevenard}\ \emph {et~al.}(2016)\citenamefont
		{Thevenard}, \citenamefont {Camara}, \citenamefont {Majrab}, \citenamefont
		{Bernard}, \citenamefont {Rovillain}, \citenamefont {Lema{\^i}tre},
		\citenamefont {Gourdon},\ and\ \citenamefont
		{Duquesne}}]{thevenard2016precessional}%
	\BibitemOpen
	\bibfield  {author} {\bibinfo {author} {\bibfnamefont {L.}~\bibnamefont
			{Thevenard}}, \bibinfo {author} {\bibfnamefont {I.~S.}\ \bibnamefont
			{Camara}}, \bibinfo {author} {\bibfnamefont {S.}~\bibnamefont {Majrab}},
		\bibinfo {author} {\bibfnamefont {M.}~\bibnamefont {Bernard}}, \bibinfo
		{author} {\bibfnamefont {P.}~\bibnamefont {Rovillain}}, \bibinfo {author}
		{\bibfnamefont {A.}~\bibnamefont {Lema{\^i}tre}}, \bibinfo {author}
		{\bibfnamefont {C.}~\bibnamefont {Gourdon}},\ and\ \bibinfo {author}
		{\bibfnamefont {J.-Y.}\ \bibnamefont {Duquesne}},\ }\bibfield  {title}
	{\bibinfo {title} {Precessional magnetization switching by a surface acoustic
			wave},\ }\href {https://doi.org/10.1103/PhysRevB.93.134430} {\bibfield
		{journal} {\bibinfo  {journal} {Physical Review B}\ }\textbf {\bibinfo
			{volume} {93}},\ \bibinfo {pages} {134430} (\bibinfo {year}
		{2016})}\BibitemShut {NoStop}%
	\bibitem [{\citenamefont {Labanowski}\ \emph {et~al.}(2016)\citenamefont
		{Labanowski}, \citenamefont {Jung},\ and\ \citenamefont
		{Salahuddin}}]{Labanowski2016}%
	\BibitemOpen
	\bibfield  {author} {\bibinfo {author} {\bibfnamefont {D.}~\bibnamefont
			{Labanowski}}, \bibinfo {author} {\bibfnamefont {A.}~\bibnamefont {Jung}},\
		and\ \bibinfo {author} {\bibfnamefont {S.}~\bibnamefont {Salahuddin}},\
	}\bibfield  {title} {\bibinfo {title} {Power absorption in acoustically
			driven ferromagnetic resonance},\ }\href {https://doi.org/10.1063/1.4939914}
	{\bibfield  {journal} {\bibinfo  {journal} {Appl. Phys. Lett.}\ }\textbf
		{\bibinfo {volume} {108}},\ \bibinfo {pages} {022905} (\bibinfo {year}
		{2016})}\BibitemShut {NoStop}%
	\bibitem [{\citenamefont {Foerster}\ \emph {et~al.}(2017)\citenamefont
		{Foerster}, \citenamefont {Maci{\`a}}, \citenamefont {Statuto}, \citenamefont
		{Finizio}, \citenamefont {Hern{\'a}ndez-M{\'i}nguez}, \citenamefont
		{Lend{\'i}nez}, \citenamefont {Santos}, \citenamefont {Fontcuberta},
		\citenamefont {Hern{\`a}ndez}, \citenamefont {Kl{\"a}ui} \emph
		{et~al.}}]{foerster2017direct}%
	\BibitemOpen
	\bibfield  {author} {\bibinfo {author} {\bibfnamefont {M.}~\bibnamefont
			{Foerster}}, \bibinfo {author} {\bibfnamefont {F.}~\bibnamefont {Maci{\`a}}},
		\bibinfo {author} {\bibfnamefont {N.}~\bibnamefont {Statuto}}, \bibinfo
		{author} {\bibfnamefont {S.}~\bibnamefont {Finizio}}, \bibinfo {author}
		{\bibfnamefont {A.}~\bibnamefont {Hern{\'a}ndez-M{\'i}nguez}}, \bibinfo
		{author} {\bibfnamefont {S.}~\bibnamefont {Lend{\'i}nez}}, \bibinfo {author}
		{\bibfnamefont {P.~V.}\ \bibnamefont {Santos}}, \bibinfo {author}
		{\bibfnamefont {J.}~\bibnamefont {Fontcuberta}}, \bibinfo {author}
		{\bibfnamefont {J.~M.}\ \bibnamefont {Hern{\`a}ndez}}, \bibinfo {author}
		{\bibfnamefont {M.}~\bibnamefont {Kl{\"a}ui}}, \emph {et~al.},\ }\bibfield
	{title} {\bibinfo {title} {Direct imaging of delayed magneto-dynamic modes
			induced by surface acoustic waves},\ }\href
	{https://doi.org/10.1038/s41467-017-00456-0} {\bibfield  {journal} {\bibinfo
			{journal} {Nat. Commun.}\ }\textbf {\bibinfo {volume} {8}},\ \bibinfo {pages}
		{407} (\bibinfo {year} {2017})}\BibitemShut {NoStop}%
	\bibitem [{\citenamefont {Kuszewski}\ \emph {et~al.}(2018)\citenamefont
		{Kuszewski}, \citenamefont {Camara}, \citenamefont {Biarrotte}, \citenamefont
		{Becerra}, \citenamefont {von Bardeleben}, \citenamefont {Torres},
		\citenamefont {Lema{\^{\i}}tre}, \citenamefont {Gourdon}, \citenamefont
		{Duquesne},\ and\ \citenamefont {Thevenard}}]{Kuszewski_2018}%
	\BibitemOpen
	\bibfield  {author} {\bibinfo {author} {\bibfnamefont {P.}~\bibnamefont
			{Kuszewski}}, \bibinfo {author} {\bibfnamefont {I.~S.}\ \bibnamefont
			{Camara}}, \bibinfo {author} {\bibfnamefont {N.}~\bibnamefont {Biarrotte}},
		\bibinfo {author} {\bibfnamefont {L.}~\bibnamefont {Becerra}}, \bibinfo
		{author} {\bibfnamefont {J.}~\bibnamefont {von Bardeleben}}, \bibinfo
		{author} {\bibfnamefont {W.~S.}\ \bibnamefont {Torres}}, \bibinfo {author}
		{\bibfnamefont {A.}~\bibnamefont {Lema{\^{\i}}tre}}, \bibinfo {author}
		{\bibfnamefont {C.}~\bibnamefont {Gourdon}}, \bibinfo {author} {\bibfnamefont
			{J.-Y.}\ \bibnamefont {Duquesne}},\ and\ \bibinfo {author} {\bibfnamefont
			{L.}~\bibnamefont {Thevenard}},\ }\bibfield  {title} {\bibinfo {title}
		{Resonant magneto-acoustic switching: influence of rayleigh wave frequency
			and wavevector},\ }\href {https://doi.org/10.1088/1361-648x/aac152}
	{\bibfield  {journal} {\bibinfo  {journal} {J. Phys. Condens. Matter}\
		}\textbf {\bibinfo {volume} {30}},\ \bibinfo {pages} {244003} (\bibinfo
		{year} {2018})}\BibitemShut {NoStop}%
	\bibitem [{\citenamefont {Foerster}\ \emph {et~al.}(2018)\citenamefont
		{Foerster}, \citenamefont {Aballe}, \citenamefont {HernÃ ndez},\ and\
		\citenamefont {MaciÃ }}]{MRS2018}%
	\BibitemOpen
	\bibfield  {author} {\bibinfo {author} {\bibfnamefont {M.}~\bibnamefont
			{Foerster}}, \bibinfo {author} {\bibfnamefont {L.}~\bibnamefont {Aballe}},
		\bibinfo {author} {\bibfnamefont {J.~M.}\ \bibnamefont {HernÃ ndez}},\ and\
		\bibinfo {author} {\bibfnamefont {F.}~\bibnamefont {MaciÃ }},\ }\bibfield
	{title} {\bibinfo {title} {Subnanosecond magnetization dynamics driven by
			strain waves},\ }\href {https://doi.org/10.1557/mrs.2018.258} {\bibfield
		{journal} {\bibinfo  {journal} {MRS Bull.}\ }\textbf {\bibinfo {volume}
			{43}},\ \bibinfo {pages} {854} (\bibinfo {year} {2018})}\BibitemShut
	{NoStop}%
	\bibitem [{\citenamefont {Adhikari}\ and\ \citenamefont
		{Adenwalla}(2021)}]{adhikari2021surface}%
	\BibitemOpen
	\bibfield  {author} {\bibinfo {author} {\bibfnamefont {A.}~\bibnamefont
			{Adhikari}}\ and\ \bibinfo {author} {\bibfnamefont {S.}~\bibnamefont
			{Adenwalla}},\ }\bibfield  {title} {\bibinfo {title} {Surface acoustic waves
			increase magnetic domain wall velocity},\ }\href
	{https://doi.org/10.1063/9.0000159} {\bibfield  {journal} {\bibinfo
			{journal} {AIP Advances}\ }\textbf {\bibinfo {volume} {11}},\ \bibinfo
		{pages} {015234} (\bibinfo {year} {2021})}\BibitemShut {NoStop}%
	\bibitem [{\citenamefont {Casals}\ \emph {et~al.}(2020)\citenamefont {Casals},
		\citenamefont {Statuto}, \citenamefont {Foerster}, \citenamefont
		{Hern{\'a}ndez-M{\'i}nguez}, \citenamefont {Cichelero}, \citenamefont
		{Manshausen}, \citenamefont {Mandziak}, \citenamefont {Aballe}, \citenamefont
		{Hern{\`a}ndez},\ and\ \citenamefont {Maci{\`a}}}]{casals2020generation}%
	\BibitemOpen
	\bibfield  {author} {\bibinfo {author} {\bibfnamefont {B.}~\bibnamefont
			{Casals}}, \bibinfo {author} {\bibfnamefont {N.}~\bibnamefont {Statuto}},
		\bibinfo {author} {\bibfnamefont {M.}~\bibnamefont {Foerster}}, \bibinfo
		{author} {\bibfnamefont {A.}~\bibnamefont {Hern{\'a}ndez-M{\'i}nguez}},
		\bibinfo {author} {\bibfnamefont {R.}~\bibnamefont {Cichelero}}, \bibinfo
		{author} {\bibfnamefont {P.}~\bibnamefont {Manshausen}}, \bibinfo {author}
		{\bibfnamefont {A.}~\bibnamefont {Mandziak}}, \bibinfo {author}
		{\bibfnamefont {L.}~\bibnamefont {Aballe}}, \bibinfo {author} {\bibfnamefont
			{J.~M.}\ \bibnamefont {Hern{\`a}ndez}},\ and\ \bibinfo {author}
		{\bibfnamefont {F.}~\bibnamefont {Maci{\`a}}},\ }\bibfield  {title} {\bibinfo
		{title} {Generation and imaging of magnetoacoustic waves over millimeter
			distances},\ }\href {https://doi.org/10.1103/PhysRevLett.124.137202}
	{\bibfield  {journal} {\bibinfo  {journal} {Physical Review Letters}\
		}\textbf {\bibinfo {volume} {124}},\ \bibinfo {pages} {137202} (\bibinfo
		{year} {2020})}\BibitemShut {NoStop}%
	\bibitem [{\citenamefont {M{\"u}ller}\ \emph {et~al.}(2022)\citenamefont
		{M{\"u}ller}, \citenamefont {Durdaut}, \citenamefont {Holl{\"a}nder},
		\citenamefont {Kittmann}, \citenamefont {Schell}, \citenamefont {Meyners},
		\citenamefont {H{\"o}ft}, \citenamefont {Quandt},\ and\ \citenamefont
		{McCord}}]{mueller2022imaging}%
	\BibitemOpen
	\bibfield  {author} {\bibinfo {author} {\bibfnamefont {C.}~\bibnamefont
			{M{\"u}ller}}, \bibinfo {author} {\bibfnamefont {P.}~\bibnamefont {Durdaut}},
		\bibinfo {author} {\bibfnamefont {R.~B.}\ \bibnamefont {Holl{\"a}nder}},
		\bibinfo {author} {\bibfnamefont {A.}~\bibnamefont {Kittmann}}, \bibinfo
		{author} {\bibfnamefont {V.}~\bibnamefont {Schell}}, \bibinfo {author}
		{\bibfnamefont {D.}~\bibnamefont {Meyners}}, \bibinfo {author} {\bibfnamefont
			{M.}~\bibnamefont {H{\"o}ft}}, \bibinfo {author} {\bibfnamefont
			{E.}~\bibnamefont {Quandt}},\ and\ \bibinfo {author} {\bibfnamefont
			{J.}~\bibnamefont {McCord}},\ }\bibfield  {title} {\bibinfo {title} {Imaging
			of love waves and their interaction with magnetic domain walls in
			magnetoelectric magnetic field sensors},\ }\href
	{https://doi.org/10.1002/aelm.202200033} {\bibfield  {journal} {\bibinfo
			{journal} {Advanced Electronic Materials}\ }\textbf {\bibinfo {volume} {8}},\
		\bibinfo {pages} {2200033} (\bibinfo {year} {2022})}\BibitemShut {NoStop}%
	\bibitem [{\citenamefont {Seemann}\ \emph {et~al.}(2022)\citenamefont
		{Seemann}, \citenamefont {Gomonay}, \citenamefont {Mokrousov}, \citenamefont
		{H{\"o}rner}, \citenamefont {Valencia}, \citenamefont {Klamser},
		\citenamefont {Kronast}, \citenamefont {Erb}, \citenamefont {Hindmarch},
		\citenamefont {Wixforth} \emph {et~al.}}]{seemann2022magnetoelastic}%
	\BibitemOpen
	\bibfield  {author} {\bibinfo {author} {\bibfnamefont {K.~M.}\ \bibnamefont
			{Seemann}}, \bibinfo {author} {\bibfnamefont {O.}~\bibnamefont {Gomonay}},
		\bibinfo {author} {\bibfnamefont {Y.}~\bibnamefont {Mokrousov}}, \bibinfo
		{author} {\bibfnamefont {A.}~\bibnamefont {H{\"o}rner}}, \bibinfo {author}
		{\bibfnamefont {S.}~\bibnamefont {Valencia}}, \bibinfo {author}
		{\bibfnamefont {P.}~\bibnamefont {Klamser}}, \bibinfo {author} {\bibfnamefont
			{F.}~\bibnamefont {Kronast}}, \bibinfo {author} {\bibfnamefont
			{A.}~\bibnamefont {Erb}}, \bibinfo {author} {\bibfnamefont {A.}~\bibnamefont
			{Hindmarch}}, \bibinfo {author} {\bibfnamefont {A.}~\bibnamefont {Wixforth}},
		\emph {et~al.},\ }\bibfield  {title} {\bibinfo {title} {Magnetoelastic
			resonance as a probe for exchange springs at antiferromagnet-ferromagnet
			interfaces},\ }\href {https://doi.org/10.1103/PhysRevB.105.144432} {\bibfield
		{journal} {\bibinfo  {journal} {Physical Review B}\ }\textbf {\bibinfo
			{volume} {105}},\ \bibinfo {pages} {144432} (\bibinfo {year}
		{2022})}\BibitemShut {NoStop}%
	\bibitem [{\citenamefont {Delsing}\ \emph {et~al.}(2019)\citenamefont
		{Delsing}, \citenamefont {Cleland}, \citenamefont {Schuetz}, \citenamefont
		{KnÃ¶rzer}, \citenamefont {Giedke}, \citenamefont {Cirac}, \citenamefont
		{Srinivasan}, \citenamefont {Wu} \emph {et~al.}}]{SAW_roadmap_2019}%
	\BibitemOpen
	\bibfield  {author} {\bibinfo {author} {\bibfnamefont {P.}~\bibnamefont
			{Delsing}}, \bibinfo {author} {\bibfnamefont {A.~N.}\ \bibnamefont
			{Cleland}}, \bibinfo {author} {\bibfnamefont {M.~J.~A.}\ \bibnamefont
			{Schuetz}}, \bibinfo {author} {\bibfnamefont {J.}~\bibnamefont {KnÃ¶rzer}},
		\bibinfo {author} {\bibfnamefont {G.}~\bibnamefont {Giedke}}, \bibinfo
		{author} {\bibfnamefont {J.~I.}\ \bibnamefont {Cirac}}, \bibinfo {author}
		{\bibfnamefont {K.}~\bibnamefont {Srinivasan}}, \bibinfo {author}
		{\bibfnamefont {M.}~\bibnamefont {Wu}}, \emph {et~al.},\ }\bibfield  {title}
	{\bibinfo {title} {The 2019 surface acoustic waves roadmap},\ }\href
	{https://doi.org/10.1088/1361-6463/ab1b04} {\bibfield  {journal} {\bibinfo
			{journal} {Journal of Physics D: Applied Physics}\ }\textbf {\bibinfo
			{volume} {52}},\ \bibinfo {pages} {353001} (\bibinfo {year}
		{2019})}\BibitemShut {NoStop}%
	\bibitem [{\citenamefont {Yang}\ and\ \citenamefont
		{Schmidt}(2021)}]{yang2021acoustic}%
	\BibitemOpen
	\bibfield  {author} {\bibinfo {author} {\bibfnamefont {W.-G.}\ \bibnamefont
			{Yang}}\ and\ \bibinfo {author} {\bibfnamefont {H.}~\bibnamefont {Schmidt}},\
	}\bibfield  {title} {\bibinfo {title} {Acoustic control of magnetism toward
			energy-efficient applications},\ }\href {https://doi.org/10.1063/5.0042138}
	{\bibfield  {journal} {\bibinfo  {journal} {Applied Physics Reviews}\
		}\textbf {\bibinfo {volume} {8}},\ \bibinfo {pages} {021304} (\bibinfo {year}
		{2021})}\BibitemShut {NoStop}%
	\bibitem [{\citenamefont {Puebla}\ \emph {et~al.}(2022)\citenamefont {Puebla},
		\citenamefont {Hwang}, \citenamefont {Maekawa},\ and\ \citenamefont
		{Otani}}]{puebla2022perspectives}%
	\BibitemOpen
	\bibfield  {author} {\bibinfo {author} {\bibfnamefont {J.}~\bibnamefont
			{Puebla}}, \bibinfo {author} {\bibfnamefont {Y.}~\bibnamefont {Hwang}},
		\bibinfo {author} {\bibfnamefont {S.}~\bibnamefont {Maekawa}},\ and\ \bibinfo
		{author} {\bibfnamefont {Y.}~\bibnamefont {Otani}},\ }\bibfield  {title}
	{\bibinfo {title} {Perspectives on spintronics with surface acoustic waves},\
	}\href {https://doi.org/10.1063/5.0093654} {\bibfield  {journal} {\bibinfo
			{journal} {Appl. Phys. Lett.}\ }\textbf {\bibinfo {volume} {120}},\ \bibinfo
		{pages} {220502} (\bibinfo {year} {2022})}\BibitemShut {NoStop}%
	\bibitem [{\citenamefont {Dong}\ and\ \citenamefont
		{Zaghloul}(2019)}]{dong2019generation}%
	\BibitemOpen
	\bibfield  {author} {\bibinfo {author} {\bibfnamefont {B.}~\bibnamefont
			{Dong}}\ and\ \bibinfo {author} {\bibfnamefont {M.~E.}\ \bibnamefont
			{Zaghloul}},\ }\bibfield  {title} {\bibinfo {title} {Generation and
			enhancement of surface acoustic waves on highly doped p-type $\text{GaAs}$
			substrate},\ }\href {https://doi.org/10.1039/C9NA00281B} {\bibfield
		{journal} {\bibinfo  {journal} {Nanoscale Advances}\ }\textbf {\bibinfo
			{volume} {1}},\ \bibinfo {pages} {3537} (\bibinfo {year} {2019})}\BibitemShut
	{NoStop}%
	\bibitem [{\citenamefont {Rampal}\ and\ \citenamefont
		{Kleiman}(2021)}]{rampal2021optical}%
	\BibitemOpen
	\bibfield  {author} {\bibinfo {author} {\bibfnamefont {A.}~\bibnamefont
			{Rampal}}\ and\ \bibinfo {author} {\bibfnamefont {R.~N.}\ \bibnamefont
			{Kleiman}},\ }\bibfield  {title} {\bibinfo {title} {Optical actuation of a
			micromechanical photodiode via the photovoltaic-piezoelectric effect},\
	}\href {https://doi.org/10.1038/s41378-021-00249-y} {\bibfield  {journal}
		{\bibinfo  {journal} {Microsystems \& Nanoengineering}\ }\textbf {\bibinfo
			{volume} {7}},\ \bibinfo {pages} {29} (\bibinfo {year} {2021})}\BibitemShut
	{NoStop}%
	\bibitem [{\citenamefont {Aballe}\ \emph {et~al.}(2015)\citenamefont {Aballe},
		\citenamefont {Foerster}, \citenamefont {Pellegrin}, \citenamefont
		{Nicolas},\ and\ \citenamefont {Ferrer}}]{aballe2015alba}%
	\BibitemOpen
	\bibfield  {author} {\bibinfo {author} {\bibfnamefont {L.}~\bibnamefont
			{Aballe}}, \bibinfo {author} {\bibfnamefont {M.}~\bibnamefont {Foerster}},
		\bibinfo {author} {\bibfnamefont {E.}~\bibnamefont {Pellegrin}}, \bibinfo
		{author} {\bibfnamefont {J.}~\bibnamefont {Nicolas}},\ and\ \bibinfo {author}
		{\bibfnamefont {S.}~\bibnamefont {Ferrer}},\ }\bibfield  {title} {\bibinfo
		{title} {The $\text{ALBA}$ spectroscopic $\text{LEEM-PEEM}$ experimental
			station: layout and performance},\ }\href
	{https://doi.org/10.1107/S1600577515003537} {\bibfield  {journal} {\bibinfo
			{journal} {J. Sync. Rad.}\ }\textbf {\bibinfo {volume} {22}},\ \bibinfo
		{pages} {745} (\bibinfo {year} {2015})}\BibitemShut {NoStop}%
	\bibitem [{\citenamefont {Wadley}\ \emph {et~al.}(2013)\citenamefont {Wadley},
		\citenamefont {Nov{\'a}k}, \citenamefont {Campion}, \citenamefont {Rinaldi},
		\citenamefont {Marti}, \citenamefont {Reichlov{\'a}}, \citenamefont
		{Å½elezn{\'y}}, \citenamefont {G{\'a}zquez}, \citenamefont {Roldan},
		\citenamefont {Varela}, \citenamefont {Khalyavin}, \citenamefont {Langridge},
		\citenamefont {Kriegner}, \citenamefont {M{\'a}ca}, \citenamefont {Masek},
		\citenamefont {Bertacco}, \citenamefont {Hol{\'y}}, \citenamefont
		{Rushforth}, \citenamefont {Edmonds}, \citenamefont {Gallagher},
		\citenamefont {Foxon}, \citenamefont {Wunderlich},\ and\ \citenamefont
		{Jungwirth}}]{Wadley2013}%
	\BibitemOpen
	\bibfield  {author} {\bibinfo {author} {\bibfnamefont {P.}~\bibnamefont
			{Wadley}}, \bibinfo {author} {\bibfnamefont {V.}~\bibnamefont {Nov{\'a}k}},
		\bibinfo {author} {\bibfnamefont {R.~P.}\ \bibnamefont {Campion}}, \bibinfo
		{author} {\bibfnamefont {C.}~\bibnamefont {Rinaldi}}, \bibinfo {author}
		{\bibfnamefont {X.}~\bibnamefont {Marti}}, \bibinfo {author} {\bibfnamefont
			{H.}~\bibnamefont {Reichlov{\'a}}}, \bibinfo {author} {\bibfnamefont
			{J.}~\bibnamefont {Å½elezn{\'y}}}, \bibinfo {author} {\bibfnamefont
			{J.}~\bibnamefont {G{\'a}zquez}}, \bibinfo {author} {\bibfnamefont {M.~A.}\
			\bibnamefont {Roldan}}, \bibinfo {author} {\bibfnamefont {M.}~\bibnamefont
			{Varela}}, \bibinfo {author} {\bibfnamefont {D.~D.}\ \bibnamefont
			{Khalyavin}}, \bibinfo {author} {\bibfnamefont {S.}~\bibnamefont
			{Langridge}}, \bibinfo {author} {\bibfnamefont {D.}~\bibnamefont {Kriegner}},
		\bibinfo {author} {\bibfnamefont {F.}~\bibnamefont {M{\'a}ca}}, \bibinfo
		{author} {\bibfnamefont {J.~D.~I.}\ \bibnamefont {Masek}}, \bibinfo {author}
		{\bibfnamefont {R.}~\bibnamefont {Bertacco}}, \bibinfo {author}
		{\bibfnamefont {V.}~\bibnamefont {Hol{\'y}}}, \bibinfo {author}
		{\bibfnamefont {A.~W.}\ \bibnamefont {Rushforth}}, \bibinfo {author}
		{\bibfnamefont {K.~W.}\ \bibnamefont {Edmonds}}, \bibinfo {author}
		{\bibfnamefont {B.~L.}\ \bibnamefont {Gallagher}}, \bibinfo {author}
		{\bibfnamefont {C.~T.}\ \bibnamefont {Foxon}}, \bibinfo {author}
		{\bibfnamefont {J.}~\bibnamefont {Wunderlich}},\ and\ \bibinfo {author}
		{\bibfnamefont {T.}~\bibnamefont {Jungwirth}},\ }\bibfield  {title} {\bibinfo
		{title} {Tetragonal phase of epitaxial room-temperature antiferromagnet
			$\text{CuMnAs}$},\ }\href {https://doi.org/10.1038/ncomms3322} {\bibfield
		{journal} {\bibinfo  {journal} {Nature Communications}\ }\textbf {\bibinfo
			{volume} {4}},\ \bibinfo {pages} {2322} (\bibinfo {year} {2013})}\BibitemShut
	{NoStop}%
	\bibitem [{\citenamefont {von Boehn}\ \emph {et~al.}(2020)\citenamefont {von
			Boehn}, \citenamefont {Foerster}, \citenamefont {von Boehn}, \citenamefont
		{Prat}, \citenamefont {Maci{\`a}}, \citenamefont {Casals}, \citenamefont
		{Khaliq}, \citenamefont {Hern{\'a}ndez-M{\'i}nguez}, \citenamefont {Aballe},\
		and\ \citenamefont {Imbihl}}]{von2020promotion}%
	\BibitemOpen
	\bibfield  {author} {\bibinfo {author} {\bibfnamefont {B.}~\bibnamefont {von
				Boehn}}, \bibinfo {author} {\bibfnamefont {M.}~\bibnamefont {Foerster}},
		\bibinfo {author} {\bibfnamefont {M.}~\bibnamefont {von Boehn}}, \bibinfo
		{author} {\bibfnamefont {J.}~\bibnamefont {Prat}}, \bibinfo {author}
		{\bibfnamefont {F.}~\bibnamefont {Maci{\`a}}}, \bibinfo {author}
		{\bibfnamefont {B.}~\bibnamefont {Casals}}, \bibinfo {author} {\bibfnamefont
			{M.}~\bibnamefont {Khaliq}}, \bibinfo {author} {\bibfnamefont
			{A.}~\bibnamefont {Hern{\'a}ndez-M{\'i}nguez}}, \bibinfo {author}
		{\bibfnamefont {L.}~\bibnamefont {Aballe}},\ and\ \bibinfo {author}
		{\bibfnamefont {R.}~\bibnamefont {Imbihl}},\ }\bibfield  {title} {\bibinfo
		{title} {On the promotion of catalytic reactions by surface acoustic waves},\
	}\href {https://doi.org/10.1002/anie.202005883} {\bibfield  {journal}
		{\bibinfo  {journal} {Angew. Chem. Int. Ed.}\ }\textbf {\bibinfo {volume}
			{59}},\ \bibinfo {pages} {20224} (\bibinfo {year} {2020})}\BibitemShut
	{NoStop}%
	\bibitem [{\citenamefont {Nateprov}\ \emph {et~al.}(2011)\citenamefont
		{Nateprov}, \citenamefont {Kravtsov}, \citenamefont {Fritsch},\ and\
		\citenamefont {von L{\"o}hneysen}}]{nateprov2011structure}%
	\BibitemOpen
	\bibfield  {author} {\bibinfo {author} {\bibfnamefont {A.~N.}\ \bibnamefont
			{Nateprov}}, \bibinfo {author} {\bibfnamefont {V.~C.}\ \bibnamefont
			{Kravtsov}}, \bibinfo {author} {\bibfnamefont {V.}~\bibnamefont {Fritsch}},\
		and\ \bibinfo {author} {\bibfnamefont {H.}~\bibnamefont {von
				L{\"o}hneysen}},\ }\bibfield  {title} {\bibinfo {title} {Structure and
			properties of the tetragonal phase of $\text{MnCuAs}$},\ }\href
	{https://doi.org/10.3103/S1068375511060147} {\bibfield  {journal} {\bibinfo
			{journal} {Surf. Engin. Appl. Electrochem.}\ }\textbf {\bibinfo {volume}
			{47}},\ \bibinfo {pages} {540} (\bibinfo {year} {2011})}\BibitemShut
	{NoStop}%
	\bibitem [{\citenamefont {Hills}\ \emph {et~al.}(2015)\citenamefont {Hills},
		\citenamefont {Wadley}, \citenamefont {Campion}, \citenamefont {Beardsley},
		\citenamefont {Edmonds}, \citenamefont {Gallagher}, \citenamefont {Novak},
		\citenamefont {Ouladdiaf},\ and\ \citenamefont
		{Jungwirth}}]{hills2015paramagnetic}%
	\BibitemOpen
	\bibfield  {author} {\bibinfo {author} {\bibfnamefont {V.}~\bibnamefont
			{Hills}}, \bibinfo {author} {\bibfnamefont {P.}~\bibnamefont {Wadley}},
		\bibinfo {author} {\bibfnamefont {R.~P.}\ \bibnamefont {Campion}}, \bibinfo
		{author} {\bibfnamefont {R.}~\bibnamefont {Beardsley}}, \bibinfo {author}
		{\bibfnamefont {K.~W.}\ \bibnamefont {Edmonds}}, \bibinfo {author}
		{\bibfnamefont {B.~L.}\ \bibnamefont {Gallagher}}, \bibinfo {author}
		{\bibfnamefont {V.}~\bibnamefont {Novak}}, \bibinfo {author} {\bibfnamefont
			{B.}~\bibnamefont {Ouladdiaf}},\ and\ \bibinfo {author} {\bibfnamefont
			{T.}~\bibnamefont {Jungwirth}},\ }\bibfield  {title} {\bibinfo {title}
		{Paramagnetic to antiferromagnetic transition in epitaxial tetragonal
			$\text{CuMnAs}$},\ }\href {https://doi.org/10.1063/1.4914119} {\bibfield
		{journal} {\bibinfo  {journal} {Journal of Applied Physics}\ }\textbf
		{\bibinfo {volume} {117}},\ \bibinfo {pages} {172608} (\bibinfo {year}
		{2015})}\BibitemShut {NoStop}%
	\bibitem [{\citenamefont {Wadley}\ \emph {et~al.}(2015)\citenamefont {Wadley},
		\citenamefont {Hills}, \citenamefont {Shahedkhah}, \citenamefont {Edmonds},
		\citenamefont {Campion}, \citenamefont {Nov{\'a}k}, \citenamefont
		{Ouladdiaf}, \citenamefont {Khalyavin}, \citenamefont {Langridge},
		\citenamefont {Saidl}, \citenamefont {Nemec}, \citenamefont {Rushforth},
		\citenamefont {Gallagher}, \citenamefont {Dhesi}, \citenamefont
		{Maccherozzi}, \citenamefont {{\v{Z}}elezn{\'y}},\ and\ \citenamefont
		{Jungwirth}}]{wadley2015antiferromagnetic}%
	\BibitemOpen
	\bibfield  {author} {\bibinfo {author} {\bibfnamefont {P.}~\bibnamefont
			{Wadley}}, \bibinfo {author} {\bibfnamefont {V.}~\bibnamefont {Hills}},
		\bibinfo {author} {\bibfnamefont {M.}~\bibnamefont {Shahedkhah}}, \bibinfo
		{author} {\bibfnamefont {K.}~\bibnamefont {Edmonds}}, \bibinfo {author}
		{\bibfnamefont {R.}~\bibnamefont {Campion}}, \bibinfo {author} {\bibfnamefont
			{V.}~\bibnamefont {Nov{\'a}k}}, \bibinfo {author} {\bibfnamefont
			{B.}~\bibnamefont {Ouladdiaf}}, \bibinfo {author} {\bibfnamefont
			{D.}~\bibnamefont {Khalyavin}}, \bibinfo {author} {\bibfnamefont
			{S.}~\bibnamefont {Langridge}}, \bibinfo {author} {\bibfnamefont
			{V.}~\bibnamefont {Saidl}}, \bibinfo {author} {\bibfnamefont
			{P.}~\bibnamefont {Nemec}}, \bibinfo {author} {\bibfnamefont
			{A.}~\bibnamefont {Rushforth}}, \bibinfo {author} {\bibfnamefont
			{B.}~\bibnamefont {Gallagher}}, \bibinfo {author} {\bibfnamefont
			{S.}~\bibnamefont {Dhesi}}, \bibinfo {author} {\bibfnamefont
			{F.}~\bibnamefont {Maccherozzi}}, \bibinfo {author} {\bibfnamefont
			{J.}~\bibnamefont {{\v{Z}}elezn{\'y}}},\ and\ \bibinfo {author}
		{\bibfnamefont {T.}~\bibnamefont {Jungwirth}},\ }\bibfield  {title} {\bibinfo
		{title} {Antiferromagnetic structure in tetragonal $\text{CuMnAs}$ thin
			films},\ }\href {https://doi.org/10.1038/srep17079} {\bibfield  {journal}
		{\bibinfo  {journal} {Scientific Reports}\ }\textbf {\bibinfo {volume} {5}},\
		\bibinfo {pages} {17079} (\bibinfo {year} {2015})}\BibitemShut {NoStop}%
	\bibitem [{\citenamefont {Rovirola}\ \emph {et~al.}(2022)\citenamefont
		{Rovirola}, \citenamefont {Khaliq}, \citenamefont {Casals}, \citenamefont
		{Foerster}, \citenamefont {NiÃ±o}, \citenamefont {Herfort}, \citenamefont
		{HernÃ ndez}, \citenamefont {MaciÃ },\ and\ \citenamefont
		{HernÃ¡ndez-MÃ­nguez}}]{rovirola2022resonant}%
	\BibitemOpen
	\bibfield  {author} {\bibinfo {author} {\bibfnamefont {M.}~\bibnamefont
			{Rovirola}}, \bibinfo {author} {\bibfnamefont {M.~W.}\ \bibnamefont
			{Khaliq}}, \bibinfo {author} {\bibfnamefont {B.}~\bibnamefont {Casals}},
		\bibinfo {author} {\bibfnamefont {M.}~\bibnamefont {Foerster}}, \bibinfo
		{author} {\bibfnamefont {M.~A.}\ \bibnamefont {NiÃ±o}}, \bibinfo {author}
		{\bibfnamefont {J.}~\bibnamefont {Herfort}}, \bibinfo {author} {\bibfnamefont
			{J.~M.}\ \bibnamefont {HernÃ ndez}}, \bibinfo {author} {\bibfnamefont
			{F.}~\bibnamefont {MaciÃ }},\ and\ \bibinfo {author} {\bibfnamefont
			{A.}~\bibnamefont {HernÃ¡ndez-MÃ­nguez}},\ }\href@noop {} {\bibinfo {title}
		{Resonant and off-resonant magnetoacoustic waves in epitaxial
			{{Fe$_3$Si/GaAs}} hybrid structures}} (\bibinfo {year} {2022}),\ \Eprint
	{https://arxiv.org/abs/2212.07994} {arXiv:2212.07994 [cond-mat.mtrl-sci]}
	\BibitemShut {NoStop}%
	\bibitem [{\citenamefont {Grigorev}\ \emph {et~al.}(2022)\citenamefont
		{Grigorev}, \citenamefont {Filianina}, \citenamefont {Lytvynenko},
		\citenamefont {Sobolev}, \citenamefont {Pokharel}, \citenamefont {Sapozhnik},
		\citenamefont {Kleibert}, \citenamefont {Bodnar}, \citenamefont {Grigorev},
		\citenamefont {Skourski}, \citenamefont {KlÃ¤ui}, \citenamefont {Elmers},
		\citenamefont {Jourdan},\ and\ \citenamefont
		{Demsar}}]{grigorev2022opticallytriggered}%
	\BibitemOpen
	\bibfield  {author} {\bibinfo {author} {\bibfnamefont {V.}~\bibnamefont
			{Grigorev}}, \bibinfo {author} {\bibfnamefont {M.}~\bibnamefont {Filianina}},
		\bibinfo {author} {\bibfnamefont {Y.}~\bibnamefont {Lytvynenko}}, \bibinfo
		{author} {\bibfnamefont {S.}~\bibnamefont {Sobolev}}, \bibinfo {author}
		{\bibfnamefont {A.~R.}\ \bibnamefont {Pokharel}}, \bibinfo {author}
		{\bibfnamefont {A.}~\bibnamefont {Sapozhnik}}, \bibinfo {author}
		{\bibfnamefont {A.}~\bibnamefont {Kleibert}}, \bibinfo {author}
		{\bibfnamefont {S.~Y.}\ \bibnamefont {Bodnar}}, \bibinfo {author}
		{\bibfnamefont {P.}~\bibnamefont {Grigorev}}, \bibinfo {author}
		{\bibfnamefont {Y.}~\bibnamefont {Skourski}}, \bibinfo {author}
		{\bibfnamefont {M.}~\bibnamefont {KlÃ¤ui}}, \bibinfo {author} {\bibfnamefont
			{H.-J.}\ \bibnamefont {Elmers}}, \bibinfo {author} {\bibfnamefont
			{M.}~\bibnamefont {Jourdan}},\ and\ \bibinfo {author} {\bibfnamefont
			{J.}~\bibnamefont {Demsar}},\ }\href@noop {} {\bibinfo {title}
		{Optically-triggered strain-driven {N\'{e}el} vector manipulation in a
			metallic antiferromagnet}} (\bibinfo {year} {2022}),\ \Eprint
	{https://arxiv.org/abs/2205.05411} {arXiv:2205.05411 [cond-mat.mtrl-sci]}
	\BibitemShut {NoStop}%
	\bibitem [{\citenamefont {Lyons}\ \emph {et~al.}(2023)\citenamefont {Lyons},
		\citenamefont {Puebla}, \citenamefont {Yamamoto}, \citenamefont {Deacon},
		\citenamefont {Hwang}, \citenamefont {Ishibashi}, \citenamefont {Maekawa},\
		and\ \citenamefont {Otani}}]{lyons2023acoustically}%
	\BibitemOpen
	\bibfield  {author} {\bibinfo {author} {\bibfnamefont {T.~P.}\ \bibnamefont
			{Lyons}}, \bibinfo {author} {\bibfnamefont {J.}~\bibnamefont {Puebla}},
		\bibinfo {author} {\bibfnamefont {K.}~\bibnamefont {Yamamoto}}, \bibinfo
		{author} {\bibfnamefont {R.~S.}\ \bibnamefont {Deacon}}, \bibinfo {author}
		{\bibfnamefont {Y.}~\bibnamefont {Hwang}}, \bibinfo {author} {\bibfnamefont
			{K.}~\bibnamefont {Ishibashi}}, \bibinfo {author} {\bibfnamefont
			{S.}~\bibnamefont {Maekawa}},\ and\ \bibinfo {author} {\bibfnamefont
			{Y.}~\bibnamefont {Otani}},\ }\href@noop {} {\bibinfo {title} {Acoustically
			driven magnon-phonon coupling in a layered antiferromagnet}} (\bibinfo {year}
	{2023}),\ \Eprint {https://arxiv.org/abs/2303.08305} {arXiv:2303.08305
		[cond-mat.mes-hall]} \BibitemShut {NoStop}%
\end{thebibliography}

\providecommand{\noopsort}[1]{}\providecommand{\singleletter}[1]{#1}%

\end{document}